\documentclass[conference]{IEEEtran}
\IEEEoverridecommandlockouts

\usepackage[utf8]{inputenc}
\usepackage{amsmath,amsthm,amsfonts,amssymb}
\usepackage[pdftex,dvipsnames]{xcolor}
\usepackage{graphicx}
\usepackage{subcaption}
\usepackage{enumerate}
\usepackage{todonotes}
\usepackage{geometry}
\usepackage{cite}
\newgeometry{top=1in,bottom=0.75in,right=0.75in,left=0.75in}

\usepackage{algorithm}
\usepackage{algpseudocode}

\newtheorem{theorem}{Theorem} 

\newtheorem{lemma}[theorem]{Lemma}

\newtheorem{assum}{Assumption}

\def \rad{\mathrm{rad}}
\usepackage{calc}
\usepackage{tikz}
\usetikzlibrary{shapes, decorations.markings, calc}

\newcommand{\zono}[1]{\langle #1 \rangle}

\title{Resilient Set-based State Estimation for Linear Time-Invariant Systems Using Zonotopes
\thanks{This work is supported by the Swedish Research Council and the Knut and Alice Wallenberg Foundation, Sweden. It has also received funding from the European Union's Horizon Research and Innovation Programme under grant agreement No. 830927 and Marie Sk\l{}odowska-Curie grant agreement No. 101062523.}
}
\author{Muhammad Umar B. Niazi$^*$ \and Amr Alanwar$^\dagger$ \and Michelle S. Chong$^\diamond$  \and Karl Henrik Johansson$^*$
\thanks{$*$ Division of Decision and Control Systems, Digital Futures, School of Electrical Engineering and Computer Science, KTH Royal Institute of Technology, Stockholm, Sweden  (Email: \{mubniazi, kallej\}@kth.se)}
\thanks{$\dagger$ Department of Computer Science and Electrical Engineering, Jacobs University, Bremen, Germany (Email: {a.alanwar@jacobs-university.de})}
\thanks{$\diamond$ Control Systems Technology Section, Department of Mechanical Engineering, Eindhoven University of Technology, the Netherlands (Email: m.s.t.chong@tue.nl)}
}

\begin{document}

\maketitle

\begin{abstract}
    This paper considers the problem of set-based state estimation for linear time-invariant (LTI) systems under time-varying sensor attacks. Provided that the LTI system is stable and observable via every single sensor and that at least one sensor is uncompromised, we guarantee that the true state is always contained in the estimated set. We use zonotopes to represent these sets for computational efficiency. However, we show that intelligently designed stealthy attacks may cause exponential growth in the algorithm's worst-case complexity. We present several strategies to handle this complexity issue and illustrate our resilient zonotope-based state estimation algorithm on a rotating target system.
\end{abstract}

\section{Introduction}


The interconnectedness of control systems consisting of the physical system, sensors, actuators, and the communication channel that connects them, exposes its vulnerabilities to adversarial attackers in a multitude of ways. Chief among the vulnerable points of entry is via the sensors, whose measurements can be maliciously manipulated by the attacker to cause undesirable disruptions or behaviors. To ensure that the corrupted sensor data does not degrade estimation accuracy, and thereby guaranteeing reasonable control performance, several mitigation strategies have been recently proposed. Such strategies, usually known as \textit{resilient} or \textit{secure} state estimation, often leverage the redundancy of sensors (c.f. \cite{shoukry2017secure, chong2020secure, kim2018detection, pajic2016attack, he2021secure,conf:SSEreach}) to obtain estimates that converge to a neighborhood of the true state in the presence of additive sensor attacks, modulo noise and disturbances, where the estimation error bound remains uninfluenced by the attacker. 

To the best of our knowledge, most resilient state estimation techniques are point-based, where the estimated state up to any point in time is a trajectory.
However, the error bounds of point-based estimators turn out to be quite conservative, and they fail to provide precise robust guarantees.

For the state estimation of dynamical systems, stochastic filtering approaches, such as Kalman filters, assume that the statistics of the underlying random process generating the process and measurement noise is known. 
In applications where noise statistics cannot be known, such filtering approaches perform rather poorly \cite{shaked1992}. For non-stochastic uncertainties, $\mathcal{H}_\infty$ filters and observers provide a robust solution to the state estimation problem; however, they turn out to be overly conservative \cite{simon2006}. To evade these limitations and at the same time obtain precise robust guarantees, the set-based zonotopic filtering paradigm has proven to be very promising \cite{conf:AlthoffJagatjournal, conf:annualreviewAlthoff, de2022zonotopic}, with many real-world applications including fault diagnosis in industrial systems \cite{blesa2012robust}, underwater robotics \cite{conf:setmem2water}, vehicle localization \cite{conf:setloc}, and leakage detection in water distribution networks \cite{rego2021}. 
    
In safety-critical applications, guaranteed state inclusion in a bounded set is crucial to provably avoid unsafe regions in the state space. This motivates the need for set-based state estimation to obtain a set of all possible states under unknown disturbances and measurement errors belonging to known bounded sets.
In this regard, set-based estimation has a long history and was first studied by \cite{conf:setmem1971} in 1971. Recent works on the topic are \cite{conf:AlthoffJagatjournal,conf:annualreviewAlthoff} and the references therein. 

To the best of our knowledge, the literature on resilient set-based state estimation when some of the sensors are vulnerable to adversarial attacks is scarce. Only \cite{shinohara2018} and its journal version \cite{shinohara2018b} present a set-based resilient state estimation technique that relies on reachability. However, for obtaining an accurate set-based estimate, they require that the full state vector can be measured by any subset of sensors with cardinality equal to the number of safe sensors, which is quite a restrictive assumption. Without this, their guarantees on the estimation accuracy become very conservative. 

In this paper, we do not require the full state vector to be measured by any subset of sensors, but the LTI system is observable via every sensor. This assumption is required since we do not limit the number of attacked sensors to be less than half the number of sensors, and we also allow the attacker to change the set of compromised sensors at any time. Subject to these assumptions, we present a zonotope-based state estimation algorithm for LTI systems under sensor attacks. We guarantee that the true state is always included in the estimated set. A strength of our proposed scheme is that we can handle attacks compromising different sensors over time as long as at least one sensor remains untouched. Further, if the attacker compromises the same set of attacked sensors over time, we provide a detection scheme to identify the set of attacked sensors. One major drawback of the zonotope-based algorithm is its complexity, which can increase exponentially in the worst-case if stealthy attacks are employed. We discuss and demonstrate complexity reduction schemes to help with the implementation of our zonotope-based resilient state estimation algorithm.

The rest of the paper is organized as follows. We define the notations used and background needed in Section~\ref{sec:notation}. Next, the problem and assumptions are stated in Section~\ref{sec:prob}. We then present the resilient zonotopic state estimation algorithm and guarantee that the true state is always within the estimated set in Section~\ref{sec:algo}. In Section~\ref{sec:case}, we present cases where our results can be sharpened and discuss the complexity of the proposed scheme. The efficacy of the algorithm is illustrated by an example in Section~\ref{sec:evaluation}. We conclude the paper with Section~\ref{sec:conclude}.


\section{Notations and Preliminaries} \label{sec:notation}

\subsection{Notations}

The set of real numbers and integers are denoted by $\mathbb{R}$ and $\mathbb{Z}$, respectively, and $\mathbb{Z}_{\geq i} \doteq \{i,i+1,i+2,\dots\}$. A finite set of integers $\{i,i+1,i+2,\dots,i+k\}$ is denoted as $\mathbb{Z}_{[i,i+k]}$. The Euclidean norm of a vector $x \in \mathbb{R}^{n}$ is denoted as $\|x\|\doteq\sqrt{x^{T}x}$ and the maximum norm as $\|x\|_\infty\doteq \max_{i\in\{1,\dots,n\}} |x_i|$. Given a signal $v:\mathbb{Z}_{\geq 0}\rightarrow\mathbb{R}^n$, we denote its restriction to the domain $[0,k]$ by $v_{[0,k]}$, for some $k\in\mathbb{Z}_{\geq 0}$. For a set $\mathcal{S}$, $|\mathcal{S}|$ denotes its cardinality. Given sets $\mathcal{S}_1,\dots,\mathcal{S}_n$, we denote their collection as $\mathcal{S}=\{\mathcal{S}_i\}_{i\in \mathbb{Z}_{[1,n]}}$.

\subsection{Set Representations}\label{sec:zonotope}

Given a center $c_{z}\in\mathbb{R}^n$ and generator matrix $G_z\in\mathbb{R}^{n\times \xi_z}$, a \textit{zonotope} $\mathcal{Z}\subset\mathbb{R}^n$ is the set
\[
\mathcal{Z}\doteq \{c_z+G_z\beta_z : \beta_z\in[-1,1]^{\xi_z}\}
\]
where $\xi_z$ is the number of generators of $\mathcal{Z}$. Since a zonotope can be completely characterized by its center and generator matrix, the notation $\mathcal{Z}=\zono{c_z,G_z}$ is used throughout the paper for brevity.

A matrix $L\in\mathbb{R}^{n'\times n}$ multiplied with a zonotope $\mathcal{Z}$ yields $L\mathcal{Z}=\zono{Lc_z,LG_z}$. Given two zonotopes $\mathcal{Z}_1=\zono{c_{z_1},G_{z_1}}$ and $\mathcal{Z}_2=\zono{c_{z_2},G_{z_2}}$, each being a subset of $\mathbb{R}^n$, their Minkowski sum is given by
\[
\mathcal{Z}_1\oplus \mathcal{Z}_2 = \zono{c_{z_1}+c_{z_2},[\arraycolsep=2pt\begin{array}{cc} G_{z_1} & G_{z_2} \end{array}]}.
\]
A \textit{constrained zonotope} is given by
\[
\mathcal{Z} \doteq \{c_z+G_z\beta_z: \beta_z\in[-1,1]^{\xi_z}, A\beta_z = b\}
\]
where $A\in\mathbb{R}^{n \times \xi_z}$ and $b\in\mathbb{R}^n$ with $n\in\mathbb{Z}_{>0}$.
The radius of a zonotope, or a constrained zonotope, is given by
\[
\rad(\mathcal{Z}) = \min \Delta ~\text{subject to}~ \mathcal{Z}\subset\Delta\mathbb{B}^n(c_z)
\]
i.e., the radius $\Delta$ of a minimal $n$-dimensional Euclidean ball $\mathbb{B}^n(c_z)$ centered at $c_z$ and inscribing $\mathcal{Z}$.

To summarize, a zonotope is an affine transformation of a hypercube and a constrained zonotope is an affine transformation of linearly constrained hypercube.

\section{Problem Definition} \label{sec:prob}

Consider an LTI system in discrete-time
\begin{subequations}
\label{eq:system}
\begin{align}
    x(k+1) &= Ax(k) + Bu(k) + w(k)  \label{eq:system_state} \\
    y^i(k) &= C_ix(k) + v^i(k) + a^i(k); \quad k\in\mathbb{Z}_{\geq 0} \label{eq:system_output}
\end{align}
\end{subequations}
where $x(k)\in\mathbb{R}^{n_x}$ is the state, $u(k)\in\mathbb{R}^{n_u}$ is a known input that is bounded, and $y^i(k)\in\mathbb{R}^{m_i}$ is the measured output of $i$-th sensor with $i\in \mathbb{Z}_{[1,p]}$ and $p$ the total number of sensors. The vector $w(k)\in\mathcal{W}$ represents the process noise, which is bounded and assumed to be contained in the zonotope $\mathcal{W}=\zono{c_w,G_w}$, and the vector $v^i(k)\in\mathcal{V}_i$ represents the measurement noise of $i$-th sensor, which is also bounded and assumed to be contained in the zonotope $\mathcal{V}_i=\zono{c_{v^i},G_{v^i}}$ for every sensor $i$. Finally, $a^i(k)\in\mathbb{R}^{m_i}$ represents the attack signal injected by the attacker to corrupt the measurement of $i$-th sensor, and it can be arbitrary and unbounded.

\begin{assum} \label{assum:main}
\begin{enumerate}[(i)]
    \item \label{assum:main_numAttacks} 
    \textbf{Upper bound on the number of attacked sensors}: The attacker can attack up to $q<p$ number of sensors, where $q$ is known a priori. However, the exact number and set of sensors which have been attacked are unknown.
    
    \item \label{assum:main_observability}
    \textbf{Observability from each sensor}: For every $i\in\mathbb{Z}_{[1,p]}$, $(A,C_i)$ is an observable pair.
    
    \item \label{assum:main_initSet}
    \textbf{Knowledge of the initial set}: The initial state $x(0)$ is contained in a zonotope $\mathcal{X}_0=\zono{c_0,G_0}$.
    
    \item \label{assum:main_bibs}
    \textbf{Bounded input bounded state stability}: There exists $M>0$ such that $\|x(k)\|_\infty\leq M$, for any $w_{[0,k]}\in\mathcal{W}\subset\mathbb{R}^{n_x}$, and bounded $u_{[0,k]}\in\mathbb{R}^{n_u}$ with $k\in\mathbb{Z}_{\geq 1}$.
\end{enumerate}
\end{assum}

Assumption~\ref{assum:main}(\ref{assum:main_numAttacks}) is fundamental in this paper because it ensures that, at every time $k\in\mathbb{Z}_{\geq 0}$, there exists a set of uncompromised sensors $\mathsf{S}_k\subset\mathbb{Z}_{[1,p]}$ with $|\mathsf{S}_k|=p-q$ such that $a^i(k)= 0_{m_i}$ for every $i\in\mathsf{S}_k$. This, along with Assumption~\ref{assum:main}(\ref{assum:main_observability}), allows us to ensure that the true state is included inside the intersection of the estimated sets of uncompromised sensors. In addition, the assumption entails that the attacker, even though omniscient about the system dynamics and noise bounds, has limited resources at hand. We remark that this assumption is certainly  not restrictive, because it neither restricts the set of attacked sensors to be static with respect to time nor requires that $q$ is less than half the number of sensors $p$. In contrast, at any time instant, the attacker can inject arbitrary attack signals to any subset of sensors with cardinality less than or equal to $q$, where $q$ is only required to be strictly less than $p$.

Assumption~\ref{assum:main}(\ref{assum:main_observability}) is required to enable decentralized set-based operations for resilient estimation without violating the robustness guarantees. Moreover, because Assumption~\ref{assum:main}(\ref{assum:main_numAttacks}) allows the attacker to attack up to $p-1$ sensors, it is necessary that the observability is guaranteed from any sensor. 

Assumption~\ref{assum:main}(\ref{assum:main_initSet}) can be easily satisfied from the operating conditions of the system, and it is not restrictive because the size of $\mathcal{X}_0$ is not required to be small. 

Finally, Assumption~\ref{assum:main}(\ref{assum:main_bibs}) demarcates the class of systems considered in this paper and assumes bounded input bounded state (BIBS) stability, which is equivalent to saying that state matrix $A$ is Schur stable (i.e., $\rho(A)<1$). While this may appear to be restrictive in comparison to other resilient state estimation schemes for LTI systems where the $A$ matrix does not need to be Schur stable (discrete-time systems), or Hurwitz (continuous-time systems), we argue that the class of BIBS stable systems is not restrictive as it is a property all control systems strive to achieve via feedback.

Under the standing assumptions stated above, we formulate the problem statement as follows.
\subsection*{Problem Statement}
Given the model $A,C_1,\dots,C_p$, noise zonotopes $\mathcal{W}$ and $\mathcal{V}_1,\dots,\mathcal{V}_p$, output measurements $y^1(k),\dots,y^p(k)$, and the maximum number $q$ of sensors that can be attacked at any time $k$, we aim to estimate a set $\hat{\mathcal{X}}_k$ guaranteeing the inclusion 
$
    x(k)\in\hat{\mathcal{X}}_k
$
for every $k\in\mathbb{Z}_{\geq 0}$, where $x(k)$ is the true state of system \eqref{eq:system}. \hfill $\diamond$

This paper achieves the aforementioned resilient set-based state estimation problem via zonotopic filtering, which we will develop in the forthcoming sections.


\section{Resilient Zonotopic Filtering} \label{sec:algo}

In this section, we propose our main algorithm for resilient set-based state estimation, which is summarized below.

\begin{algorithm}[!]
\caption{Resilient zonotope-based state estimation}\label{alg:zono_filter}
\begin{algorithmic}[1]
\Require System matrices $A$, $B$, and $C_i$, and noise zonotopes $\mathcal{W}$ and $\mathcal{V}_{i}$, for every $i\in\mathbb{Z}_{[1,p]}$; time sequence of sensor measurements $\{y^1(k),y^2(k),\dots,y^p(k)\}_{k\in\mathbb{Z}_{\geq 0}}$.

\State Initialize: $\hat{\mathcal{X}}_{0|0}=\mathcal{X}_0$
\For {$k=1,2,3,\dots$}
\State Time update: $\hat{\mathcal{X}}_{k|k-1} = A\hat{\mathcal{X}}_{k-1|k-1} \oplus Bu(k-1) \oplus \mathcal{W}$
\State Obtain $\mathcal{Y}_k^i=\zono{c_{y^i}(k), G_{y^i}(k)}$ using \eqref{eq:meas_sets}, for every sensor $i\in\mathbb{Z}_{[1,p]}$.
\State Obtain $\mathcal{I}_k^h$ using \eqref{eq:collection_meas_sets} for every index set $\mathsf{J}_h\subset\mathbb{Z}_{[1,p]}$ with cardinality $|\mathsf{J}_h|=p-q$ and $h\in\mathbb{Z}_{[1,\eta]}$.
\State Measurement update: $\hat{\mathcal{X}}_{k|k} = \hat{\mathcal{X}}_{k|k-1} \cap \{\mathcal{I}_k^h \}_{h\in\mathbb{Z}_{[1,\eta]}}$
\EndFor
\end{algorithmic}
\end{algorithm}

\subsection{Reachable set and time update step}
\label{subset_timeupdate}

The reachable set $\mathcal{R}_k$ at time $k\in\mathbb{Z}_{\geq 0}$ is the set of states to which the system may evolve given the input $u(k-1)$, and a guarantee that the previous state is contained in $\mathcal{R}_{k-1}\ni x(k-1)$ and the process noise in $\mathcal{W}\ni w(k-1)$, i.e.,
\begin{equation} 
\label{eq:reachable_set}
    \mathcal{R}_k=A\mathcal{R}_{k-1} \oplus Bu(k-1) \oplus \mathcal{W}.
\end{equation}
By Assumption~\ref{assum:main}(\ref{assum:main_initSet}), we have $x(0)\in\mathcal{X}_0$.
Thus, by choosing $\mathcal{R}_0=\mathcal{X}_0$, the reachable set in \eqref{eq:reachable_set} can be equivalently computed as
\[
    \mathcal{R}_k = A^k \mathcal{X}_0 \oplus \sum_{j=0}^{k-1} A^{k-j-1} B u(j) \oplus \bigg( \bigoplus_{j=0}^{k-1} A^{k-j-1} \mathcal{W} \bigg)
\]
and it is guaranteed that $x(k)\in\mathcal{R}_k$ for every $k\in\mathbb{Z}_{\geq 0}$. 

Because of the open-loop computation of the reachable set, i.e., it is not corrected using the sensor measurements, it turns out to be quite conservative for larger values of time $k$.
Nonetheless, the equation \eqref{eq:reachable_set} is particularly important in the time update step, also known as the prediction step, of our proposed filtering algorithm. The \textit{time update} is given by
\begin{equation}
\label{eq:time_update}
    \hat{\mathcal{X}}_{k|k-1} = A\hat{\mathcal{X}}_{k-1|k-1} \oplus Bu(k-1) \oplus \mathcal{W}
\end{equation}
where $\mathcal{R}_{k}$ is replaced by the time update $\hat{\mathcal{X}}_{k|k-1}$ and the previous reachable set $\mathcal{R}_{k-1}$ by the previous \textit{measurement update} $\hat{\mathcal{X}}_{k-1|k-1}$. The measurement update, also known as the correction step, is described in Section~\ref{subsec_measurementUpdate}.
Notice that the attacker cannot directly influence the time update \eqref{eq:time_update}, but it can influence \eqref{eq:time_update} indirectly through the measurement update $\hat{\mathcal{X}}_{k-1|k-1}$. Thus, it is important to carefully devise the measurement update, which we do in Section~\ref{subsec_measurementUpdate}, for achieving resilience against sensor attacks.

\subsection{State space region consistent with the measurements}
\label{subsec_consistent_sets}

Before presenting the measurement update, we estimate a subset of state space that is consistent with the sensor measurements.
Given the output equation \eqref{eq:system_output}, output matrix $C_i$, and the measurement noise bound $\mathcal{V}_i$, a method to find a subset of state space consistent with the sensor~$i$'s measurement $y^i(k)$ is provided in the following lemma, which is inspired by \cite{conf:dataSetBased, alanwar2022}. To this end, we employ the singular value decomposition (SVD) of the output matrix of the $i$-th sensor
\[
C_i = \left[\arraycolsep=2pt\begin{array}{cc} 
    P^i_1 & P^i_2 
    \end{array}\right] \left[\arraycolsep=2pt\begin{array}{cc}
    \Sigma_{r_i} & 0_{r_i\times (n_x-r_i)} \\
    0_{(m_i-r_i)\times r_i} & 0_{(m_i-r_i)\times (n_x\times r_i)}
    \end{array}\right] \left[\arraycolsep=2pt\begin{array}{c}
    V_1^{i\top} \\ V_2^{i\top}
    \end{array}\right]
\]
where $\text{rank}(C_i)=r_i\leq m_i$, $P^i_1\in\mathbb{R}^{m_i\times r_i},P^i_2\in\mathbb{R}^{m_i\times (m_i-r_i)},V^i_1\in\mathbb{R}^{n_x\times r_i},V^i_2\in\mathbb{R}^{n_x\times(n_x-r_i)}$, and $\Sigma_{r_i}\in\mathbb{R}^{r_i\times r_i}$ is a positive definite diagonal matrix. Then, the pseudo-inverse of $C_i$ is given by $C_i^\dagger=V_1^i \Sigma_{r_i}^{-1} P_1^{i\top}$. If $C_i$ is full row rank, i.e., $r_i=m_i$, then $C_i^\dagger = C_i^\top (C_i C_i^\top)$.

\begin{lemma} \label{lem:measSets}
    Let Assumption~\ref{assum:main}(\ref{assum:main_observability}) and (\ref{assum:main_bibs}) hold. Then, for every $i\in\mathbb{Z}_{[1,p]}$, the state space region consistent with the measurement $y^i(k)= C_ix(k) + v^i(k) + a^i(k)$ is given by the zonotope
    \begin{subequations}
    \label{eq:meas_sets}
        \begin{align}
            &\mathcal{Y}_k^i =\zono{c_{y^i}(k), G_{y^i}(k)}, ~\text{where} \\
            &\left\{\begin{array}{rcl}
            c_{y^i}(k) &=& C_i^\dagger \big( y^i(k) - c_{v^i} \big) \\
            G_{y^i}(k) &=& [\begin{array}{cc}
            C_i^\dagger G_{v^i} & M V^i_2
            \end{array}]
            \end{array}\right.
        \end{align}
    \end{subequations}
    where $M$ is given by Assumption~\ref{assum:main}(\ref{assum:main_bibs}). 
    
    Moreover, if $i\in\mathsf{S}_k$, where $\mathsf{S}_k$ is the set of uncompromised sensors at time $k$, then $x(k)\in\mathcal{Y}_k^i$, for every $k\in\mathbb{Z}_{\geq 0}$.
\end{lemma}
\begin{IEEEproof}
    Since $(A,C_i)$ is observable for every $i\in\mathsf{S}_k$, we have that the solution to
    \begin{align*}
    y^i(k) = C_ix(k) + v^i(k) =C_i x(k) + v^i(k) \pm x(k)
    \end{align*}
    given by
    \[
    x(k) = C_i^\dagger (y^i(k) - v^i(k)) + (I_{n_x} - C_i^\dagger C_i) x(k)
    \]
    is non-trivial, where $a^i(k)=0_{m_i}$ since $i\in\mathsf{S}_k$.
    Secondly, note that
    \[
        \text{im}(V_2^i)=\text{im}(I_{n_x}-C_i^\dagger C_i)=\text{ker}(C_i).
    \]
    Therefore, $(I_{n_x} - C_i^\dagger C_i) x(k)\in\text{im}(V_2^i)$. Finally, by Assumption~\ref{assum:main}(\ref{assum:main_bibs}), $x(k)\in\zono{0,MI_{n_x}}$. Thus, it holds that
    \begin{align*}
        x(k) & \in C_i^\dagger (y^i(k) - \mathcal{V}) + (I_{n_x} - C_i^\dagger C_i) \zono{0,M I_{n_x}} \\
        & = C_i^\dagger (y^i(k) - \mathcal{V}) + V_2^i \zono{0,M I_{n_x-r_i}} \\
        & = C_i^\dagger (y^i(k) - \mathcal{V}) + \zono{0,M V_2^i } \\
        & = \mathcal{Y}_k^i
    \end{align*}
    which completes the proof.
\end{IEEEproof}

By Assumption~\ref{assum:main}(\ref{assum:main_bibs}), the state space is given by the zonotope $\zono{0,M I_{n_x}}$. Subject to this assumption, \eqref{eq:meas_sets} in the above lemma computes a subset of $\zono{0,M I_{n_x}}$ that is consistent with the sensor~$i$'s measurement. Thus, if the sensor~$i$ is unattacked at time~$k$, it is guaranteed that the true state $x(k)$ is inside the set $\mathcal{Y}_k^i$. However, the guarantee doesn't hold when $i$ is under attack at time~$k$. To verify if a subset of sensors is not attacked and can be trusted, it is necessary that the intersection of their consistent sets yields a non-empty set. This intersection will discard all the sensors whose measurements are corrupted by large attack signals. However, sensors that are injected by stealthy attack signals, i.e., signals within the noise bounds, remain undetected. Nonetheless, we can ensure that there is at least one subset of sensors with cardinality $p-q$ that is guaranteed to contain the true state $x(k)$.

\begin{theorem}
\label{thm:existence_safe_subset}
Let Assumption~\ref{assum:main} hold. Then, there exists an index set $\mathsf{J}\subset\mathbb{Z}_{[1,p]}$ with cardinality $|\mathsf{J}|=p-q$ such that $x(k)\in\mathcal{I}_k$, where $\mathcal{I}_k$ is a constrained zonotope given by
\begin{equation} \label{eq:pairwise_intersections}
\mathcal{I}_k = \bigcap\limits_{i\in\mathsf{J}} \mathcal{Y}_k^i
\end{equation}
with $\mathcal{Y}_k^i$ given in \eqref{eq:meas_sets}.
\end{theorem}
\begin{IEEEproof}
By Assumption~\ref{assum:main}(\ref{assum:main_numAttacks}), the number of uncompromised sensors $|\mathsf{S}_k|\geq p-q$, because the attacker can attack only up to $q$ sensors. Thus, there exists $\mathsf{J}\subset\mathbb{Z}_{[1,p]}$ with cardinality $|\mathsf{J}|=p-q$ containing only the uncompromised sensors, i.e., $\mathsf{J}\subseteq\mathsf{S}_k$.
Since the inclusion $x(k)\in\mathcal{Y}_k^i$ for every $i\in\mathsf{S}_k$ is guaranteed by Lemma~\ref{lem:measSets}, and there exists $\mathsf{J}$ with cardinality $|\mathsf{J}|=p-q$ such that $\mathsf{J}\subseteq\mathsf{S}_k$, the inclusion $x(k)\in\mathcal{I}_k$ is guaranteed with $\mathcal{I}_k$ given in \eqref{eq:pairwise_intersections}.
\end{IEEEproof}

We have shown that there exists a subset of sensors whose consistent sets yield a non-empty intersection, and the intersection contains the true state. However, in the presence of stealthy attacks, it is not possible to completely discard the attacked sensors. There could be multiple subsets of sensors whose consistent sets yield non-empty intersections, but only some of them may contain the true state.

\subsection{Measurement update step} \label{subsec_measurementUpdate}

Measurement update $\hat{\mathcal{X}}_{k|k}$ corrects the conservative estimate of the model-based time update by incorporating new information from the sensor measurements \eqref{eq:meas_sets}.
In other words, the measurement update step involves intersecting the time update set $\hat{\mathcal{X}}_{k|k-1}$ with the state space regions consistent with the sensor measurements. 

\subsubsection{Measurement update in the absence of attacks}

First, consider the following result in the absence of the attacker.

\begin{lemma} \label{lem:no_attack}
Let Assumption~\ref{assum:main} hold with the number of attacks $q=0$ for every $k\in\mathbb{Z}_{\geq 0}$, i.e., $a^i\equiv 0_{m_i}$ for every $i\in\mathbb{Z}_{[1,p]}$. Then, for every $k\in\mathbb{Z}_{\geq 1}$, it holds that $x(k)\in\hat{\mathcal{X}}_{k|k}$, where $\hat{\mathcal{X}}_{k|k}$ is a constrained zonotope given by
\begin{align} \label{eq:xxy}
\hat{\mathcal{X}}_{k|k} = \hat{\mathcal{X}}_{k|k-1} \cap  \bigg( \bigcap_{i=1}^p \mathcal{Y}_k^i \bigg)
\end{align}
and $\hat{\mathcal{X}}_{k|k-1}$ is given in \eqref{eq:time_update} with $\hat{\mathcal{X}}_{1|0}=A\mathcal{X}_0\oplus B u(0) \oplus\mathcal{W}$.
\end{lemma}
\begin{IEEEproof}
Since $x(0)\in\mathcal{X}_0$ by Assumption~\ref{assum:main}(\ref{assum:main_initSet}), we have that $x(1)\in\hat{\mathcal{X}}_{1|0}$. Also, by Lemma~\ref{lem:measSets}, $x(1)\in\mathcal{Y}_1^i$ for every $i\in\mathbb{Z}_{[1,p]}$. Therefore, we have $x(1)\in\hat{\mathcal{X}}_{1|1}$. This, in turn, implies that $x(2)\in\hat{\mathcal{X}}_{2|1}$. By applying Lemma~\ref{lem:measSets} again, we have that $x(2)\in\hat{\mathcal{X}}_{2|2}$. Thus, by induction, for every $k\in\mathbb{Z}_{\geq 1}$, $x(k-1)\in\hat{\mathcal{X}}_{k-1|k-1}$ implies $x(k)\in\hat{\mathcal{X}}_{k|k-1}$, which guarantees $x(k)\in\hat{\mathcal{X}}_{k|k}$ by Lemma~\ref{lem:measSets}.
\end{IEEEproof}

The equation \eqref{eq:xxy} is the usual measurement update in the absence of attacker, which is central to zonotopic filtering \cite{de2022zonotopic}. However, this measurement update may yield an empty estimated set $\hat{\mathcal{X}}_{k|k}=\emptyset$ even when only one sensor is under attack. In this case, the attacker has to only ensure that the attack signal is large enough so that the consistent sets yield an empty intersection. Therefore, when considering that a subset of sensors might be attacked, a more sophisticated way of performing measurement update is developed next.

\subsubsection{Measurement update in the presence of attacks}
To obtain the measurement update $\hat{\mathcal{X}}_{k|k}$ in the presence of attacker, we propose to intersect the time update $\hat{\mathcal{X}}_{k|k-1}$ with the state space regions consistent with the measurements of all subsets of sensors with cardinality $p-q$. That is, for every index set $\mathsf{J}_h\subset\mathbb{Z}_{[1,p]}$ with cardinality $|\mathsf{J}_h|=p-q$, compute the intersection of consistent sets
\begin{equation}
\label{eq:collection_meas_sets}
    \mathcal{I}_k^h = \bigcap\limits_{j\in\mathsf{J}_h} \mathcal{Y}_k^j
    \end{equation}
where $h=1,\dots,\eta$ with
\begin{equation}
\label{eq:eta_num_intersections}
    \eta = \binom{p}{p-q} = \frac{p!}{q! (p-q)!}.
\end{equation}
Then, the measurement update $\hat{\mathcal{X}}_{k|k}$ is obtained as
\begin{equation}
\label{eq:measurement_update}
    \hat{\mathcal{X}}_{k|k} = \hat{\mathcal{X}}_{k|k-1} \cap \{\mathcal{I}_k^h \}_{h\in \mathbb{Z}_{[1,\eta]}}
\end{equation}
where we note that $\hat{\mathcal{X}}_{k|k}$ is a collection of multiple constrained zonotopes.

\begin{theorem} \label{thm:x_include}
Let Assumption~\ref{assum:main} hold. Then, given the measurement update $\hat{\mathcal{X}}_{k|k}$ from \eqref{eq:measurement_update}, the inclusion $x(k)\in\hat{\mathcal{X}}_{k|k}$ is guaranteed for every $k\in\mathbb{Z}_{\geq 1}$.
\end{theorem}
\begin{IEEEproof}[Proof idea]
The inclusion can be guaranteed through the same arguments as in the proof of Lemma~\ref{lem:no_attack} but using Theorem~\ref{thm:existence_safe_subset} instead of Lemma~\ref{lem:measSets}.
\end{IEEEproof}

Although the inclusion of the true state is guaranteed by the above theorem, it is important to remark that the number of sets in the measurement update \eqref{eq:measurement_update} may increase with respect to time under stealthy attacks. We address this issue in Section~\ref{subsec_complexity} by proposing several techniques that facilitate computational efficiency of the algorithm.

It is worth mentioning that the proposed algorithm is resilient because the attacker cannot deteriorate the estimation accuracy over time. If a subset $\mathsf{J}_h$ contains a sensor which is injected by a large attack signal, it will be automatically discarded because of an empty intersection $\mathcal{I}_k^h$ in \eqref{eq:collection_meas_sets}. Therefore, in order to yield a non-empty intersection, the attacker can only inject small attack signals whose magnitude is within the measurement noise bounds $\mathcal{V}_i$, which does not deteriorate the estimation accuracy.

\subsection{Bound on the estimation error}
\label{sec:estimation_error}

Since Theorem~\ref{thm:x_include} guarantees that the true state $x(k)$ of system~\eqref{eq:system} lies in at least one of the zonotopes in the measurement update $\hat{\mathcal{X}}_{k|k}$ at each $k\in\mathbb{Z}_{\geq 0}$, it must also lie in a zonotope that overbounds $\hat{\mathcal{X}}_{k|k}$. That is, we overbound the collection of constrained zonotopes in $\hat{\mathcal{X}}_{k|k}$ by another constrained zonotope $\hat{\mathcal{Z}}_k=\zono{\hat{c}_z(k),\hat{G}_z(k)}$, which is obtained by solving
\begin{equation} \label{eq:zono_inclusion}
\min \text{rad}(\hat{\mathcal{Z}}_k) ~\text{subject to}~ \hat{\mathcal{X}}_{k|k} \subset \hat{\mathcal{Z}}_k.
\end{equation}
Then, the estimation error can be bounded by
\[
\|\hat{c}_z(k) - x(k)\| \leq \text{rad}(\hat{\mathcal{Z}}_k).
\]
It can be proven that the error computed above is upper bounded asymptotically because of the stable time-update step (Section~\ref{subset_timeupdate}). Moreover, in practice, this error bound is significantly smaller than the error bounds obtained by point-based resilient estimators \cite{pajic2016attack, he2021secure}.

\subsection{Methods to reduce the complexity}
\label{subsec_complexity}

The major computational challenge  of Algorithm~\ref{alg:zono_filter} that can be exploited by the attacker lies in the measurement update step \eqref{eq:measurement_update} for computing $\hat{\mathcal{X}}_{k|k}$, which is a collection of zonotopes whose cardinality (i.e., the number of zonotopes) could grow over time. To reduce computational complexity resulting from the increasing cardinality of the measurement update, we propose several pruning methods. The first step is to remove the empty sets or subsets of other sets in the measurement update intersection \eqref{eq:measurement_update}. It is also possible to obtain a single overbounding zonotope of $\hat{\mathcal{X}}_{k|k}$ as in \eqref{eq:zono_inclusion}, and use it in the next time update step. However, a better trade-off between accuracy and complexity is to not overbound the whole collection, but only the intersecting zonotopes in the collection $\hat{\mathcal{X}}_{k|k}$. This may not make the cardinality of $\hat{\mathcal{X}}_{k|k}$ equal to one, but it reduces it significantly by allowing minimal loss of accuracy.

Another method is employ a point-based resilient estimator, if it exists, in parallel with the set-based resilient estimator. In this case, we may consider only those candidates in the measurement update collection that lie within the intersection of $\hat{\mathcal{X}}_{k|k}$ and an error margin generated by a point-based resilient state estimator. However, the existing point-based resilient state estimators \cite{shoukry2017secure, chong2020secure, kim2018detection, pajic2016attack, he2021secure} require that the total number of sensors be strictly greater than twice the number of compromised sensors $q<p/2$ and the members of the attacked sensors also remain unchanged over time, which are tighter requirements than our standing Assumption~\ref{assum:main}(\ref{assum:main_numAttacks}). Moreover, the error margins obtained by point-based estimators are usually very conservative.

Additionally, one may also employ zonotope reduction methods \cite{yang2018comparison} to reduce the number of generators in the zonotopes, which is often increased by the Minkowski sum operation. However, this technique may result in larger radius of $\hat{\mathcal{Z}}_k$ in \eqref{eq:zono_inclusion}.


\section{Case Studies} \label{sec:case}
In this section, we discuss three scenarios to evaluate the detection mechanisms and complexity under the proposed resilient zonotope-based state estimation algorithm.

\subsection{Detection under time-invariant attacks}
A notable relaxation of the set-based state estimation scheme in this paper over other resilient schemes is Assumption~\ref{assum:main}(\ref{assum:main_numAttacks}), which allows the attacker to compromise a different set of sensors over time. However, in the case where the set of attacked sensors is time-invariant, we can detect the set of compromised sensors by identifying the $\mathcal{Y}_{k}^{i}$ in \eqref{eq:meas_sets} which do not intersect with each other or the time update set $\hat{\mathcal{X}}_{k|k}$. To be precise, under time-invariant sensor attacks, a subset of compromised sensors can be detected over time by building the following index set
\begin{multline} \label{eq:detect}
\mathcal{D}_{k}\doteq \{ i\in\mathbb{Z}_{[1,p]} : \mathcal{Y}_{k}^{i} \cap \hat{\mathcal{X}}_{k|k-1} = \emptyset \textrm{ or } \mathcal{Y}_{k}^{i} \cap \mathcal{Y}_{k}^{j} = \emptyset, \\ \text{for every}~ j\in\mathbb{Z}_{[1,p]} \}
\end{multline}
where the cardinality $|\mathcal{D}_k|$ is a non-decreasing function of time $k$. Notice that the detector \eqref{eq:detect} may not detect stealthy attacks, where the magnitude of attack signals is within the measurement noise bound. Nonetheless, in certain cases, the attacker can be detected as illustrated in Fig.~\ref{fig:p3keq23}.

\subsection{Naive attacks are discarded automatically}

\begin{figure*}[!htbp]
\vspace{-0.05cm}
    \centering
    \begin{subfigure}[h]{0.32\textwidth}
     \centering
        \includegraphics[scale=0.24]{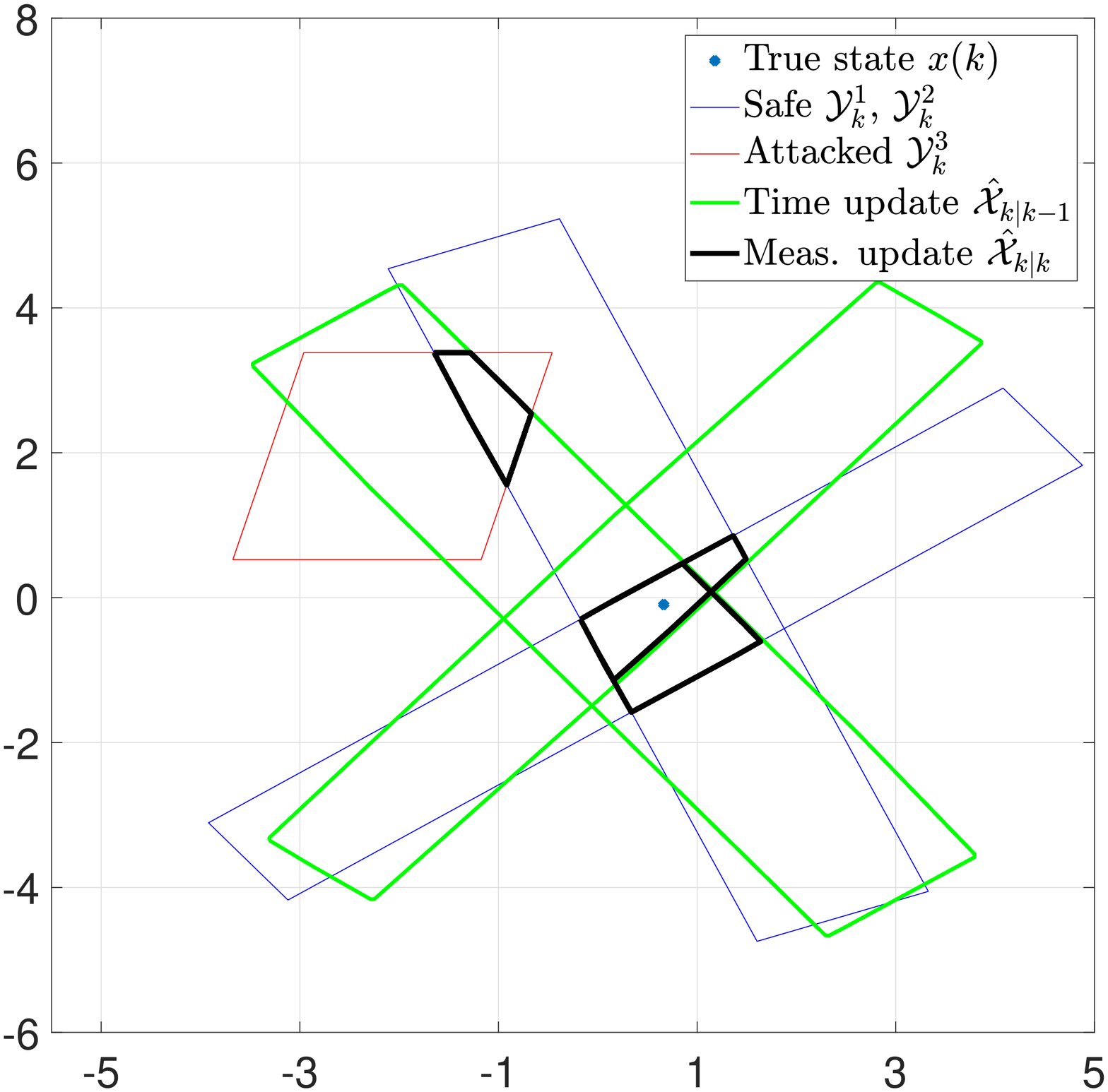}
        \caption{Time $k=4$}
        \label{fig:p1keq4}
    \end{subfigure}
    \begin{subfigure}[h]{0.32\textwidth}
     \centering
        \includegraphics[scale=0.24]{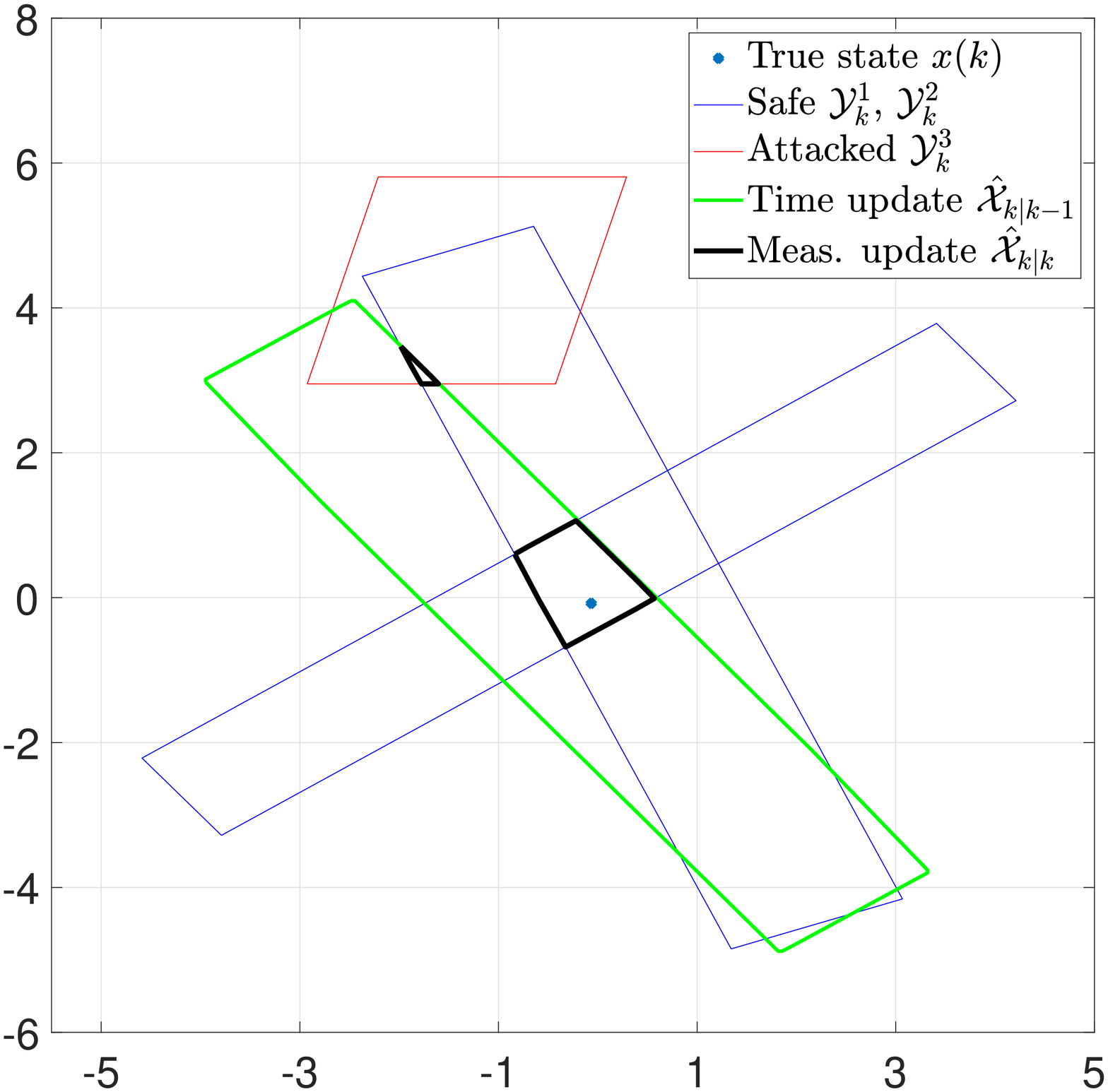}
        \caption{Time $k=12$}
        \label{fig:p2keq12}
    \end{subfigure}
    \begin{subfigure}[h]{0.32\textwidth}
     \centering
        \includegraphics[scale=0.24]{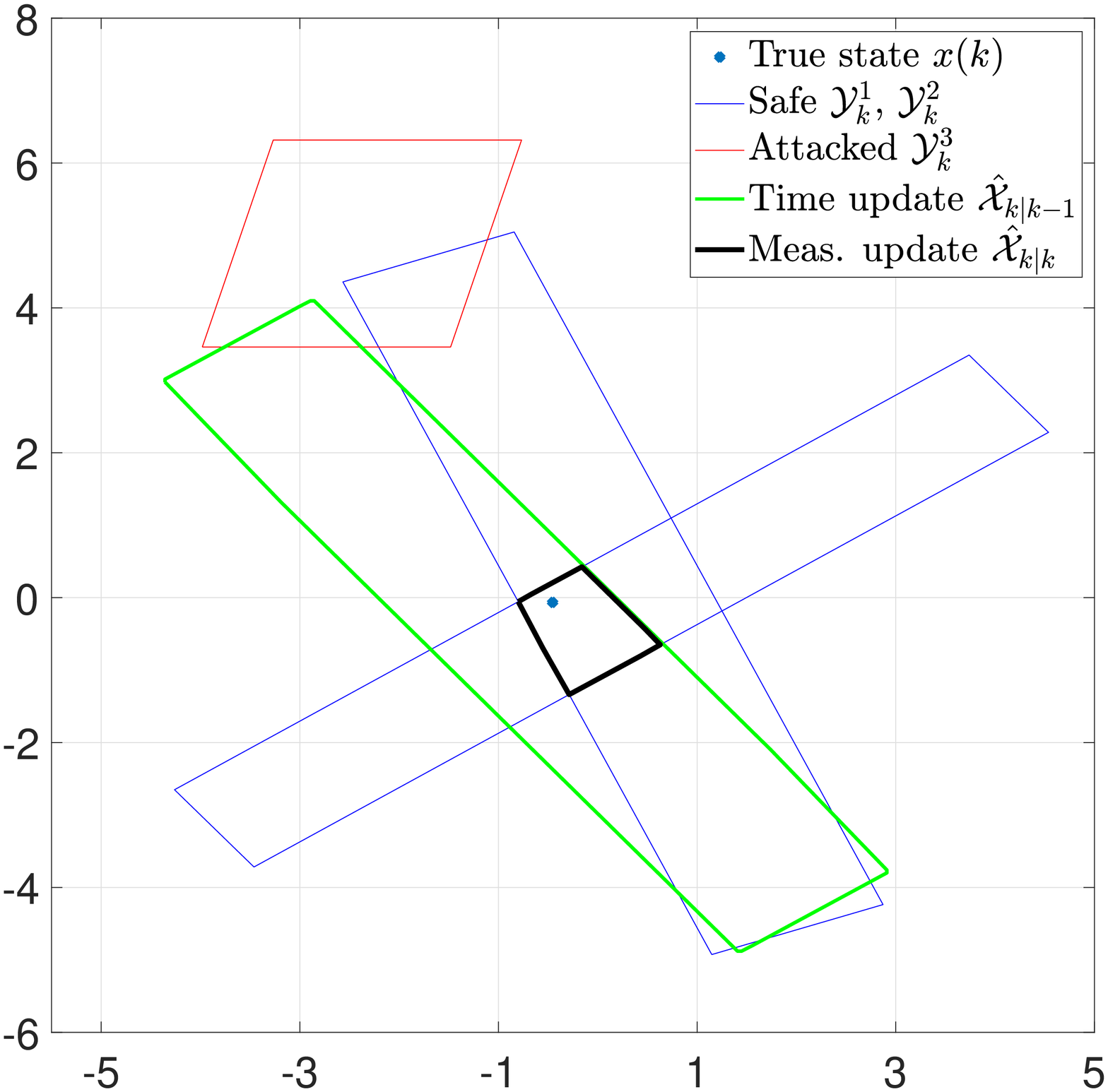}
        \caption{Time $k=23$}
        \label{fig:p3keq23}
    \end{subfigure}
\caption{Snapshots of estimated sets at different times using Algorithm~\ref{alg:zono_filter} under random attacks.}
    \label{fig:snap}
\end{figure*}

In the case where we have a naive attacker who injects large attack signals or random attack signals, the attacked sensors may be automatically discarded by our proposed Algorithm~\ref{alg:zono_filter}. As discussed already in Section~\ref{subsec_consistent_sets}, large attack signals are automatically discarded because they result in an empty intersection in \eqref{eq:collection_meas_sets}. Random attack signals, even if within the noise bounds, can also be detected eventually if the attacker is not smart enough to discount for the changing orientation of the time update set and consistent sets of other sensors. Fig.~\ref{fig:p3keq23} illustrates such a scenario.

Sensor faults like denial of service (DoS) and intermittent transmissions come under naive attacks in our proposed framework, and they can be easily handled by Algorithm~\ref{alg:zono_filter}.
Random attack signals injected by the attacker may result from their limited knowledge of the system or the noise bounds. It could also result from the fact that the attacker has limited resources at hand and cannot generate an optimal attack signal to ensure worst-case complexity at every time instant $k$. Under such assumptions, the attacked sensors can be discarded, which results in significant reduction of complexity of the measurement update step \eqref{eq:measurement_update}.

\subsection{Stealthy attacks can increase complexity exponentially in the worst-case scenario}

Stealthy attacks on sensors result in sets $\mathcal{Y}_{k}^{i}$, for $i\in\mathbb{Z}_{[1,p]}\setminus\mathsf{S}_k$, that may not yield any empty intersection in \eqref{eq:collection_meas_sets} in the worst-case scenario. This can increase the complexity of Algorithm~\ref{alg:zono_filter}, where the number of sets may increase exponentially with respect to time, which can overwhelm the available computation resources. To be precise, the number of zonotopes in the measurement update collection $\hat{\mathcal{X}}_{k|k}$ can be on the order of $\eta^k$ in the worst-case scenario, where $\eta$ is given in \eqref{eq:eta_num_intersections}. Therefore, the methods discussed in Section~\ref{subsec_complexity} are very crucial to ensure computational feasibility of the proposed algorithm at the next time instant $k+1$. Since each complexity reduction method offers a trade-off between estimation accuracy and complexity, the best method is the one that offers maximum accuracy under the available computational resources.

\begin{figure*}[!htbp]
\vspace{-0.05cm}
    \centering
    \begin{subfigure}[h]{0.32\textwidth}
     \centering
        \includegraphics[scale=0.24]{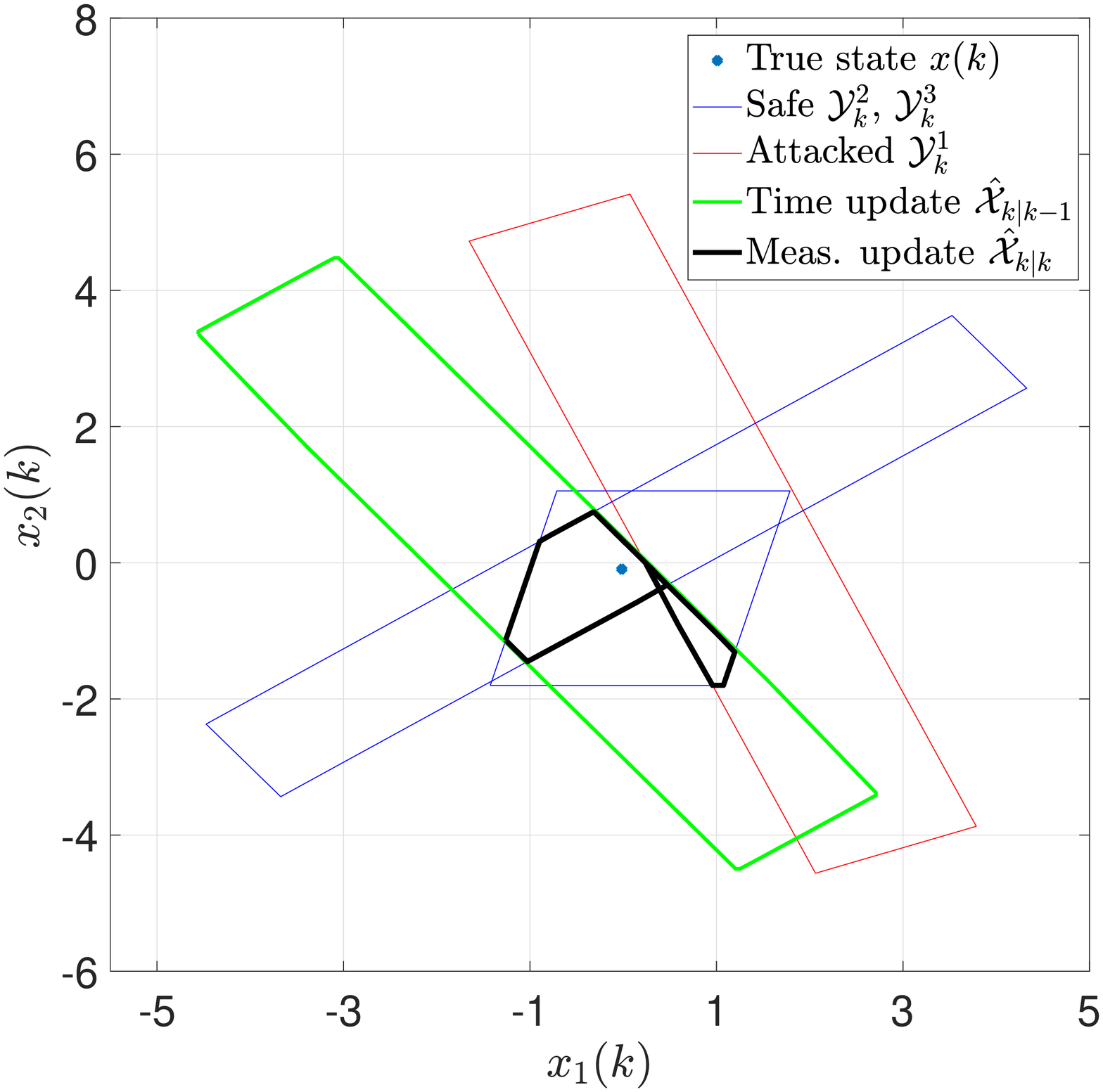}
        \caption{Time $k=3$, Sensor~1 attacked}
        \label{fig:rotAkeq3_y1attacked}
    \end{subfigure}
    \begin{subfigure}[h]{0.32\textwidth}
     \centering
        \includegraphics[scale=0.24]{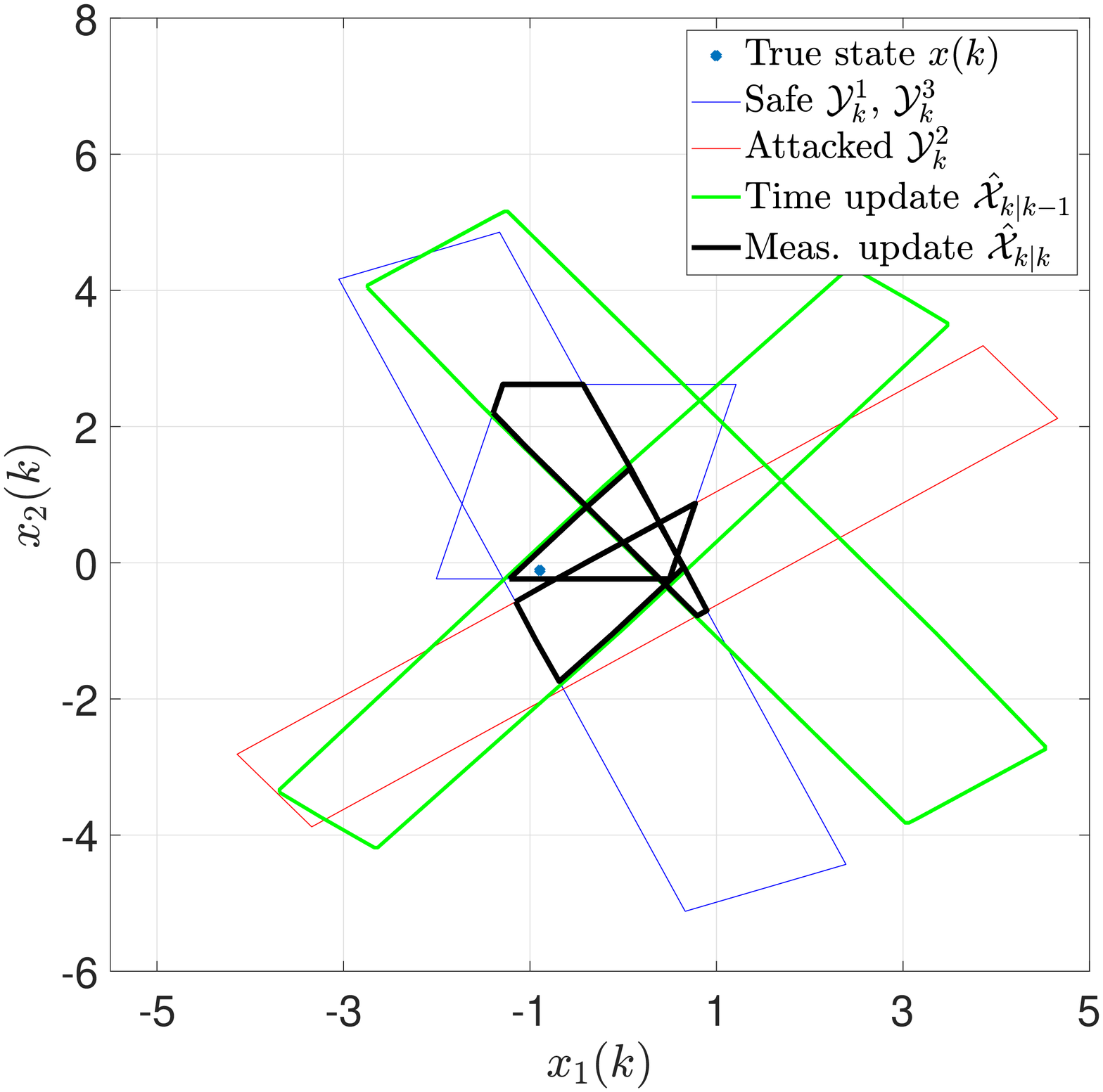}
        \caption{Time $k=4$, Sensor~2 attacked}
        \label{fig:rotAkeq4_y2attacked}
    \end{subfigure}
    \begin{subfigure}[h]{0.32\textwidth}
     \centering
        \includegraphics[scale=0.24]{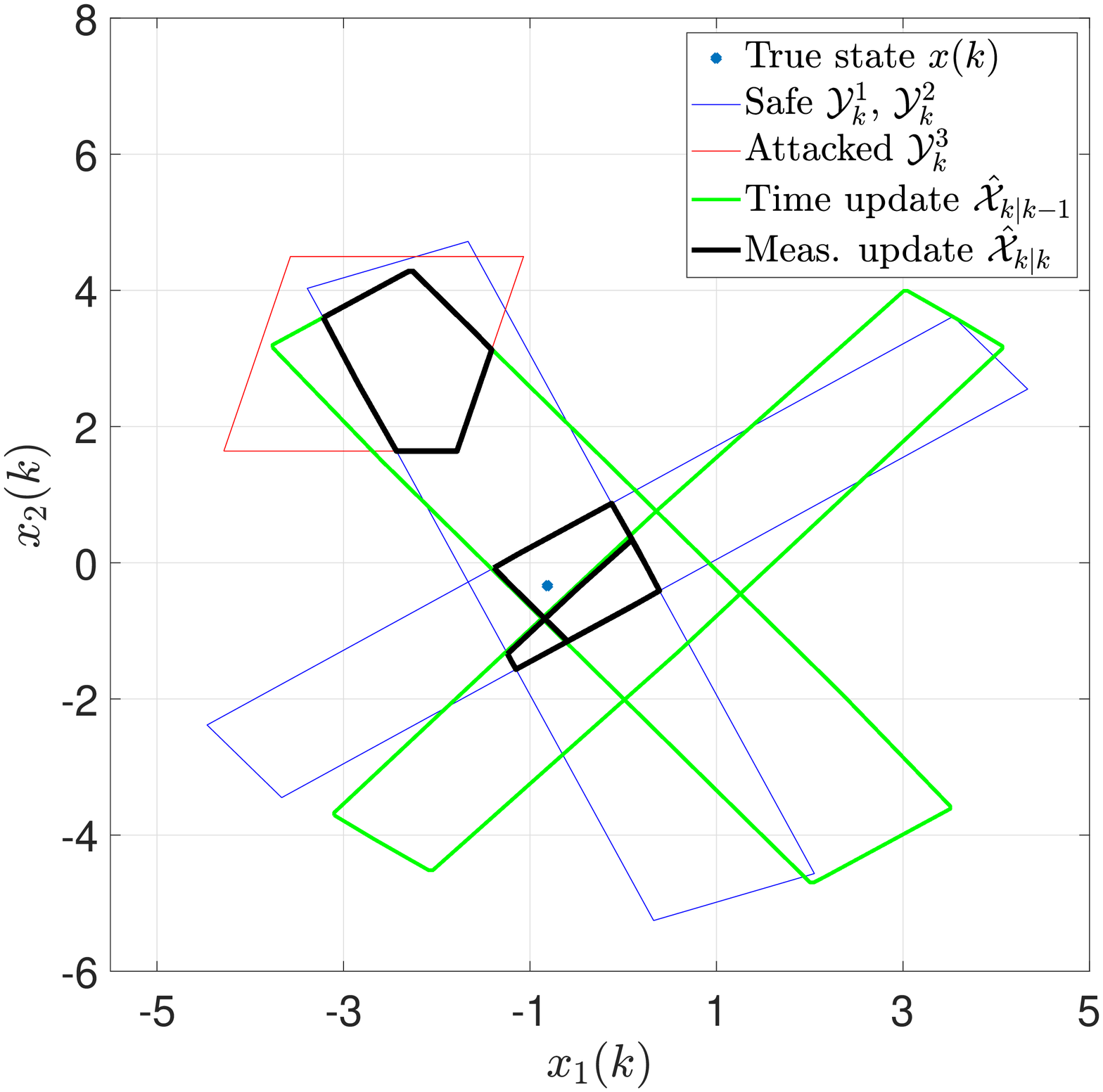}
        \caption{Time $k=5$, Sensor~3 attacked}
        \label{fig:rotAkeq5_y3attacked}
    \end{subfigure}
\caption{Snapshots of estimated sets using Algorithm~\ref{alg:zono_filter} under time-varying attack, where different sensors are attacked at different times.} 
    \label{fig:rotating_attack}
\end{figure*}

\begin{figure}[!]
    \centering
    \includegraphics[scale=0.24]{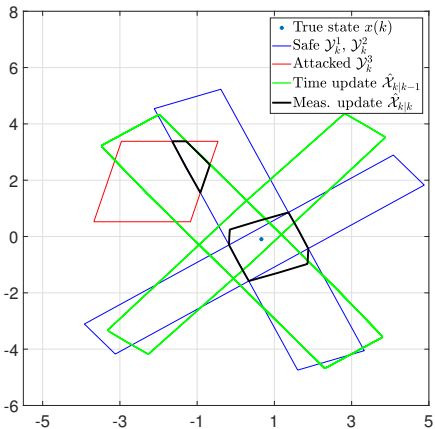}
    \caption{Over-bounding two intersecting safe sets in Fig.~\ref{fig:p2keq12} by a single set to reduce complexity.}
    \label{fig:combine_p1keq4}
\end{figure}

\section{Evaluation} 
\label{sec:evaluation}




We evaluate our method by considering an input-driven variant of the rotating target described in \cite{conf:set-diff}. The known input $u(k)\in\mathbb{R}$ is sampled uniformly from the set $\mathcal{U} = \zono{0,10}$ at every time $k$. We have
\begin{align}
    A = \begin{bmatrix}
        0.9455 & -0.2426 \\
        0.2486 & 0.9455
        \end{bmatrix},
    \quad
    B = \begin{bmatrix}
        0.1 \\ 0 
        \end{bmatrix}.
\end{align}
With the number of sensors $p=3$, we consider output matrices and respective measurement noise zonotopes as follows
\begin{align*}
    C_1 &= \begin{bmatrix}
        1 & 0.4    
        \end{bmatrix}, &
    C_2 &= \begin{bmatrix}
        0.9 & -1.2  
        \end{bmatrix}, &
    C_3 &= \begin{bmatrix}
        -0.8 & 0.2 \\ 0 & 0.7   
        \end{bmatrix} \\ 
    \mathcal{V}_{1} &= \zono{0,1}, &
    \mathcal{V}_{2} &= \zono{0,1}, &
    \mathcal{V}_{3} &= \zono{[0\;\; 0]^\top, I_2}. 
\end{align*}
The process noise signal $w(k)$ are bounded by the zonotope    $\mathcal{W} = \zono{[0\;\; 0]^\top, 0.02I_2}$. 
The noise signals $v^i(k)$ and $w(k)$ are sampled uniformly from their respective zonotope sets using the function \texttt{randPoint}$(\mathcal{Z})$ in CORA \cite{conf:cora}.

Fig.~\ref{fig:snap} presents three snapshots of the time-updated sets from the previous step (green), safe measurement consistent sets (blue), attacked measurement consistent set (red), and the final estimated measurement update sets (black). The time update sets (green) are computed using \eqref{eq:reachable_set}. Lemma~\ref{lem:measSets} is used to compute the state space regions consistent with the measurements (blue) in which one of them is under attack (red). The measurement update sets (black) are computed according to \eqref{eq:measurement_update}. It is to be noted that the true state always remains inside the measurement update sets (Theorem~\ref{thm:x_include}).

Fig.~\ref{fig:p1keq4} and \ref{fig:p2keq12} show different scenarios in which the attacked set is intersecting with the intersection of the time update set and the safe sets. On the other hand, Fig.~\ref{fig:p3keq23} shows a scenario in which the attacked set is not intersecting with the intersection of the safe and time update sets, which allows us to discard the attacked set and obtain a single measurement update set. Such a scenario may arise in non-intelligent stealthy attacks, where the attack signals are generated randomly at every time.

Fig.~\ref{fig:rotating_attack} shows a more powerful, time varying attack in which the attacker attacks a different sensor at different time steps. In Fig.~\ref{fig:rotAkeq3_y1attacked}, Sensor~$1$ is under attack and we have two estimated measurement update sets (black). Then, Sensor~$2$ is attacked in Fig.~\ref{fig:rotAkeq4_y2attacked} in which the number of estimated measurement update sets is increasing due to having a small attack value. Finally, Sensor~$3$ is attacked in Fig.~\ref{fig:rotAkeq5_y3attacked} with a larger attack value. Although the complexity increases in such attacks, it is worth noting that the true state $x(k)$ remains enclosed by the estimated measurement update sets at all time steps. Also, the estimation error remains bounded and the attacker cannot destroy the accuracy of the set-based state estimate. 

Finally, the question of reducing the complexity by minimally compromising on the accuracy remains. To this end, Fig.~\ref{fig:combine_p1keq4} illustrates one of the complexity reduction methods discussed in Section~\ref{subsec_complexity}, where we over-bound multiple intersecting sets by a single constrained zonotope. This significantly reduces the number of sets in the measurement update collection.

\section{Conclusions and Future Outlook} \label{sec:conclude}
We have presented a resilient zonotope-based state estimation scheme for LTI systems with multiple redundant sensors, i.e., the pair $(A,C_i)$ is observable from every sensor $i$. We show that our scheme ensures that the true state lies within the estimated set. We acknowledge that the scheme suffers from the curse of dimensionality and we discuss complexity reduction methods. We discuss cases under which our proposed algorithm can be sharpened, by the design of a detection algorithm in the case where the set of attacked sensors remain constant and when the attacker performs naive attacks. On the other hand, stealthy attacks can also increase the complexity of the scheme, which underlines the importance of methods for reduction.

Future work will focus on relaxing the current observable via every sensor assumption to the case where the system is observable through a subset of sensors, which will increase the applicability of the scheme.


\bibliographystyle{IEEEtran}
\bibliography{zonotopic_estimator}

\end{document}